\documentclass[
aps,
prb,
groupedaddress,
superscriptaddress,
floatfix,
twocolumn,
notitlepage
]{revtex4-1}

\usepackage{graphicx}
\usepackage{amsmath,amssymb}
\usepackage{braket}

\usepackage{hyperref} 
\usepackage{subfigure}
\usepackage{bm} 
\usepackage{mathtools} 
\usepackage{float}

\DeclareMathOperator{\sgn}{sgn}

\begin{document}

\title{Josephson Flux Flow Oscillator: the Microscopic Tunneling Approach}

\author{D. R. Gulevich} \email{d.r.gulevich@metalab.ifmo.ru}
\affiliation{ITMO University, St. Petersburg 197101, Russia}
\affiliation{Department of Physics, Loughborough University, United Kingdom}

\author{V. P. Koshelets}
\affiliation{Kotel’nikov Institute of Radio Engineering and Electronics, Russian Academy of Science, Moscow, 125009, Russia}

\author{F.\,V.~Kusmartsev}
\affiliation{Department of Physics, Loughborough University, United Kingdom}

\date{\today} 

\def\sech{{\rm sech}}
\def\rot{{\rm rot}}
\def\div{{\rm div}}
\def\arcsinh{{\rm arcsinh}}
\def\Re{{\rm Re\,}}
\def\Im{{\rm Im\,}}
\def\arccot{{\rm arccot}}
\def\p{\partial}

\begin{abstract} 
We elaborate a theoretical description of large Josephson junctions which is based on the Werthamer's microscopic tunneling theory.
The model naturally incorporates coupling of electromagnetic radiation to the tunnel currents and, therefore, is particularly suitable for description of 
the self-coupling effect in Josephson junction.
In our numerical calculations we treat the arising integro-differential equation, which describes temporal evolution of the superconducting phase difference coupled to the electromagnetic field, by the Odintsov-Semenov-Zorin algorithm.
This allows us to avoid evaluation of the time integrals at each time step while 
taking into account all the memory effects.
To validate the obtained microscopic model of large Josephson junction we focus our attention on the Josephson flux flow oscillator.
The proposed microscopic model of flux flow oscillator does not involve the phenomenological damping parameter, rather, the damping is taken into account naturally in the tunnel current amplitudes calculated at a given temperature.
The theoretically calculated current-voltage characteristics is compared to our experimental results obtained for a set of 
fabricated flux flow oscillators of different lengths.
Our theoretical calculation agrees well with the obtained experimental results, and, to our knowledge, is the first where theoretical description of Josephson flux flow oscillator is brought beyond the 
perturbed sine-Gordon equation.
\end{abstract}



{\let\newpage\relax\maketitle}

\section{Introduction}

Few years after discovery of the Josephson effect~\cite{Josephson, Anderson}
a complete microscopic description of tunnel junctions was
formulated within the tunneling Hamiltonian formalism~\cite{Cohen, AmbBar, Werthamer, Larkin}.
As a result of this effort, the microscopic tunneling theory (MTT) of Josephson tunnel junctions had emerged.
The MTT treated many of the experimentally observed tunneling phenomena fairly satisfactory, although,
few discrepancies had gradually shown up.
One of them, the famous $\cos\varphi$ problem~\cite{cos-Lang, Barone}
puzzled the scientific community for many decades.
Various experiments of the time~\cite{cos-Pedersen, cos-Falco, cos-Vincent, cos-Nisenoff, cos-Rifkin, cos-Soe} observed the sign 
of phase-dependent dissipative current, also known as the ``cosine" or quasiparticle-pair interference term, to disagree from the prediction of the MTT~\cite{Harris-1974}.
It was later suggested that, in fact, either sign is possible,
while the disagreement between the theory and experiments can be explained by broadening mechanisms which result in smearing of the Riedel peaks~\cite{Zorin, Golubov-1988}.
The MTT has also been found to overestimate the value of the critical current, which in real junctions turns out to be 
depressed by strong coupling and/or proximity effects~\cite{Broom-IBM, Broom-IEEE, Golubov-1995, Golubov-1989, Dmitriev}.
In practice, one can account for this discrepancy by a phenomenological suppression parameter~\cite{Zorin-1983}.

The MTT has been highly successful in the description of quasiparticle tunneling in superconductor-insulator-superconductor (SIS) structures and thus formed the foundations for the SIS mixer theory motivated by the unique properties offered by them in signal detection~\cite{Tucker-IEEE, Tucker-Feldman}.
Uses of the MTT include modeling SQUIDs~\cite{MTT-squid, Frank-squid}, 
Josephson arrays~\cite{Frank-array},
RSFQ logic gates and circuits~\cite{Kratz, Zorin-SFQ, OSZ, PSCAN, PSCAN-96, RSFQ-Mukhanov}.
While in the early days the attention to the phase-dependent dissipative current was motivated mainly by the $\cos\varphi$ problem, it has seen a revival very recently~\cite{cos-Lepp, cos-memristors, cos-memristors-2017, Lutchyn, Martinis, Catelani, Lenander, cos-Pop} -- this time, from the practical side: 
the  phase-dependent dissipation has found application in the proposal of superconducting memristor~\cite{cos-memristors,cos-memristors-2017},
has been considered to be a source of relaxation in superconducting qubits~\cite{Lutchyn, Martinis, Catelani, Lenander}
and even shown to be a powerful tool to suppress dissipation in fluxonium qubit~\cite{cos-Pop}.

It is, however, unfair that large Josephson junctions had been left behind in this glorious rise of the MTT.
The description of long junctions used today is still largely based on the 
sine-Gordon equation derived for tunnel junctions by Brian Josephson~\cite{Josephson-barriers}.
In the perturbed sine-Gordon equation (PSGE) used to describe large Josephson junctions,
the damping effect is usually taken into account in the form of a phenomenological ``normal" losses term proportional to the voltage~\cite{Josephson-barriers}. 
It is common in theoretical studies of large junctions to start from the PSGE as an initial point. 
To solve the PSGE several perturbative approaches had been proposed and widely used~\cite{Fogel, McLaughlin, Marcus-Imry, DGulevich-breather, nonpert, singular}.
However, note that, while the $\sin\varphi$ term describing the pair current can be justified within the MTT as a limiting case of a very slow dynamics compared to the gap frequency, the description of normal losses by the pure resistive term is rather empirical and can only be justified within a narrow temperature range close to the critical temperature~\cite{Likharev}: a condition which is rarely satisfied in real experiments. One may argue, however, that the resistive term in the PSGE is validated by the well tested, resistively and capacitively shunted junction (RCSJ) model~\cite{RCSJ-1, RCSJ-2}. The RCSJ model, however, owes its popularity to the externally shunted Josephson junctions for which it gives a quantitatively correct description at an arbitrary temperature~\cite{Likharev}. 
Obviously, this is not the case of large Josephson junctions which are rarely shunted.
Incidentally, whereas the MTT has been almost exclusively applied to small junctions, large Josephson junctions should be the first in the queue to take the cure. Owing to its naive treatment of damping, 
it is not surprising that the PSGE is not capable of reproducing essential characteristics of long Josephson junction such as subharmonic gap structures observed in experimental current-voltage characteristics (IVC).

This paper is aimed at bridging the gap between the MTT and the currently used description of large Josephson junctions. In Sec.~II we start off a revision of the MTT which we use in formulating microscopic model of 2D Josephson junction in Sec.~III.
As an application of this model, in Sec.~IV we focus our attention to the Josephson flux flow oscillator (FFO). 
To validate the developed microscopic description of FFO we compare the theoretically calculated IVCs to our experimental results for a set of FFOs of different lengths.
The last Sec.~V is devoted to discussion of a possible impact of the presented results.

\section{Review of Microscopic Tunneling Theory}

The current $I(t)$ through a Josephson junction coupled to a time-dependent electromagnetic field was calculated by Werthamer~\cite{Werthamer},
\begin{widetext}
\begin{equation}
\begin{split}
I(t)=\Im\int_{-\infty}^{\infty}\int_{-\infty}^{\infty}
d\omega\, d\omega'\; \left \{ W(\omega)W(\omega')\; e^{i(\omega+\omega' + 2 e V_{dc})t } \tilde{I}_p(\omega'+eV_{dc}) \right. 
+ \left. W(\omega)W^*(\omega')\; e^{i(\omega-\omega')t} \tilde{I}_{qp}(\omega'+eV_{dc})\right \} 
\end{split}
\label{WW}
\end{equation}
\end{widetext}
where $W(\omega)$ is defined by the time dependence of the superconducting phase difference,
\begin{equation}
\int_{-\infty}^{\infty} W(\omega) e^{i \omega t}d\omega 
= \exp\left[ \frac{i}{2}\varphi(t) - i e V_{dc} t\right],
\end{equation}
\begin{equation}
\varphi(t)  = 2 e\int^t V(t) dt,
\end{equation}
where $e>0$ is the magnitude of electron charge, $V(t)$ is the voltage across the junction and $V_{dc}$ is its DC component.
Here and in what follows we drop the Planck constant $\hbar$ where its presence is self-evident, and 
use the convention for the sign of tunnel currents as in Refs.~\onlinecite{Zorin, Tucker-IEEE, Tucker-Feldman} 
(in Refs~\onlinecite{Tucker-IEEE, Tucker-Feldman} the definition of $W(\omega)$ differs from ours by complex conjugation).
Within this convention the sign of the pair current components is chosen 
to get a positive sign in the Josephson relation $\dot\varphi=2 e V(t)$, and to restore the equality
\begin{multline}
I(t)=  \Re \tilde{I}_p(eV_{dc})\sin\varphi + \Im \tilde{I}_p (eV_{dc})\cos\varphi \\ + \Im \tilde{I}_{qp}(eV_{dc})
\label{equality}
\end{multline}
at a constant voltage.
For a symmetric junction made of identical superconductors,
the Bardeen-Cooper-Schrieffer (BCS) theory predicts singularities in the real parts (the Riedel peaks), and steps in the imaginary parts of the tunnel current amplitudes at the gap frequency $\omega_g = 2\Delta$, where $\Delta$ is the superconducting energy gap. 
The imaginary part of the quasiparticle current $\Im \tilde{I}_{qp}(eV_{dc})$ can be directly measured 
from the IVC of a voltage biased junction. There the step at the gap frequency manifests itself as a sharp rise of current at the gap voltage $V_g \equiv \omega_g/e$. 
In real systems, however, the singularities and steps are smeared by several competing effects~\cite{Zorin,Golubov-1988,Dmitriev}.

The equation~\eqref{WW} can be rewritten in the time-domain form~\cite{Harris-time-domain-1976}. 
For this, we introduce the time-domain functions $I_{p}(t)$ and $I_{qp}(t)$ 
which play a role of memory kernels and are related to the tunnel current amplitudes
$\tilde{I}_{p}(\omega)$ and $\tilde{I}_{qp}(\omega)$ by (note the difference in the sign of~$\omega$ in these two expressions~\cite{footnote1})
\begin{equation}
\begin{split}
\tilde{I}_{p}(\omega)= \int_{-\infty}^\infty I_{p}(t) e^{-i \omega t}dt
\\
\tilde{I}_{qp}(\omega)= \int_{-\infty}^\infty I_{qp}(t) e^{i \omega t}dt.
\end{split}
\label{JpJqp-kernels}
\end{equation}
The time-domain kernels in~\eqref{JpJqp-kernels} take real values and satisfy $I_{p,qp}(t)=0$ for $t<0$,
which follow from the causality properties~\cite{Harris-1975} of $\tilde{I}_{p,qp}(\omega)$.
Substituting~\eqref{JpJqp-kernels} to~\eqref{WW}, we obtain
\begin{equation}
\begin{split}
I(t) = \int_0^{\infty}\Big\{ 
I_p(t')\,\sin\left[\frac{\varphi(t)+\varphi(t-t')}{2}\right] \\ 
+\, I_{qp}(t')\,\sin
\left [ \frac{\varphi(t)-\varphi(t-t')}{2}\right ] 
\Big\}\; dt'
\end{split}
\label{J-t}
\end{equation}

Below we will be working with dimensionless units introduced as follows. The time $t$ is measured in units of the inverse of angular Josephson plasma frequency~$\omega_J$, also, we introduce
the normalized tunnel current amplitudes
\begin{equation}
\tilde{j}_{p,qp}(\xi) \equiv \frac{R_N}{V_g} \tilde{I}_{p,qp}(\xi\omega_g)
\label{j-def}
\end{equation}
where $R_N$ is the normal resistance of the junction above the gap.
From~\eqref{equality} and~\eqref{j-def} the critical current is then defined by the real part of the pair current amplitude 
at zero frequency, 
\begin{equation}
I_c=\frac{V_g}{R_N}\Re \tilde{j}_p(0).
\end{equation}
In dimensionless units the Eq.~\eqref{J-t} for the normalized current $j(t)\equiv I(t)/I_c$ takes the form
\begin{equation}
\begin{split}
j(t)=\frac{k}{\Re \tilde{j}_p(0) }\int_0^{\infty}\Big\{ j_p(kt')\,\sin\left[\frac{\varphi(t)+\varphi(t-t')}{2}\right] \\ 
+ \, j_{qp}(kt')\,\sin\left [ \frac{\varphi(t)-\varphi(t-t')}{2}\right ] \Big\} \; dt'
\end{split}
\label{j-t}
\end{equation}
where $k = \omega_g/\omega_J$ is the normalized gap frequency,
and $j_{p,qp}(\tau)$ are normalized time-domain kernels related to $\tilde{j}_{p,qp}(\xi)$ by the inverse Fourier transforms
\begin{equation}
\begin{split}
j_{p}(\tau) & = \frac{1}{2\pi} \int_{-\infty}^\infty \tilde{j}_{p}(\xi) e^{i \xi \tau}d\xi
\\
j_{qp}(\tau) & = \frac{1}{2\pi} \int_{-\infty}^\infty \tilde{j}_{qp}(\xi) e^{-i \xi \tau}d\xi.
\label{transforms}
\end{split}
\end{equation}

For the purpose of numerical calculations it is convenient to extract the normal resistance contribution from
the quasiparticle current~\cite{Harris-time-domain-1976}, introducing a {\it reduced} quasiparticle kernel $\bar{j}_{qp}(\tau)$ by setting
\begin{equation}
j_{qp}(\tau)=-\delta'(\tau-0)+\bar{j}_{qp}(\tau),
\label{reduced}
\end{equation}
the Eq.~\eqref{j-t} becomes
\begin{equation}
\begin{split}
j(t)=
\frac{k}{\Re \tilde{j}_p(0) }
\int_0^{\infty}\Big\{ 
j_p(kt')\,\sin\left[\frac{\varphi(t)+\varphi(t-t')}{2}\right] \\
+\, \bar{j}_{qp}(kt')\,\sin
\left [ \frac{\varphi(t)-\varphi(t-t')}{2}\right ] \Big\}\; dt' + \alpha_N \varphi_t
\end{split}
\label{jbar}
\end{equation}
where
\begin{equation}
\alpha_N =  \frac{1}{2k\, \Re \tilde{j}_p(0)}.
\label{alphaN}
\end{equation}
is the damping coefficient due to a pure normal resistance.

Tunnel current amplitudes $\tilde{j}_{p,qp}(\xi)$ were calculated theoretically by Werthamer~\cite{Werthamer} for zero temperature and Larkin et al.~\cite{Larkin} for arbitrary temperatures.
Unfortunately, the expressions for tunnel current amplitudes have often been given with misprints, both in the reputable sources in Josephson physics~\cite{Barone},~\cite{Likharev} and including the pioneering papers of Werthamer~\cite{Werthamer} and Larkin et al.~\cite{Larkin} themselves. For convenience, we summarize the correct expressions for tunnel current amplitudes in the Appendix and attach a summary of misprints in the existing literature
in Ref.~\onlinecite{misprints}.

The BCS theory typically predicts a higher pair current than observed experimentally (see the discussion in the Introduction). This discrepancy is taken into account by introducing a phenomenological suppression factor of the pair currents~\cite{Zorin-1983}
$\tilde{j}_p(\xi)\to \alpha_{\rm supp}\, \tilde{j}_p(\xi)$ while keeping intact the quasiparticle current. With this modification, the BCS expression for the normalized critical current is
\begin{equation}
\Re \tilde{j}_p(0) = \alpha_{\rm supp}\, \frac{\pi}{4} \tanh{\frac{\omega_g}{4k_B T}}.
\end{equation}

\section{Microscopic Tunneling Model of 2D Josephson junction}

\begin{figure}
\begin{center}
\includegraphics[width=3.5in]{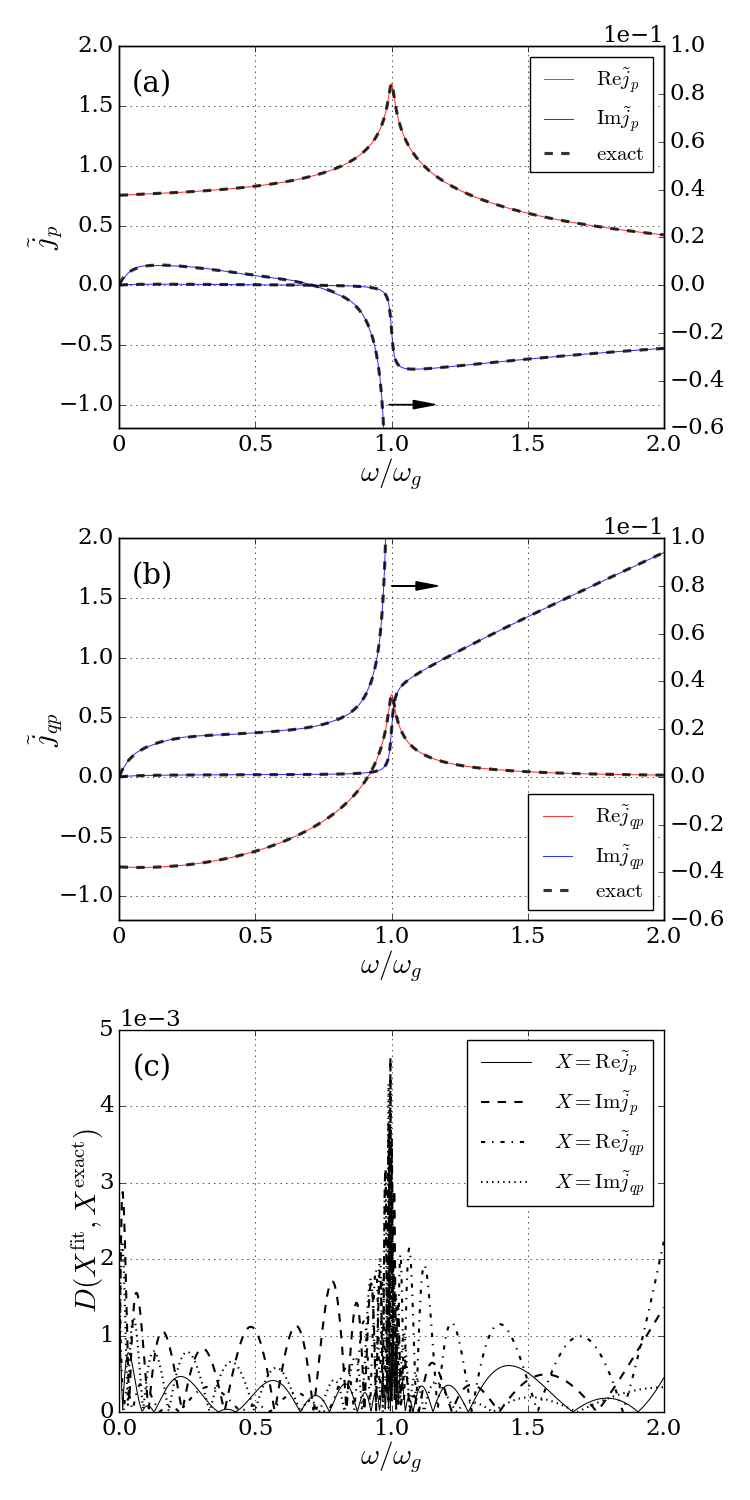}
\caption{\label{fig:amps} 
Amplitudes of the pair (a) and quasiparticle (b) tunnel currents. Solid red and blue lines represent fit to the real and imaginary parts of the pair and quasiparticle currents in the form of 
a sum of exponents~\eqref{exps} with $N=8$ terms.
The exact theoretical tunnel current amplitudes based on which the fitting was done, are shown by dashed lines for comparison.
To illustrate the behavior of the tunnel current amplitudes in the subgap region, 20x zoom of the imaginary parts of the tunnel current amplitudes is shown in both figures.
Relative difference of the fitted and exact amplitudes defined by Eq.~\eqref{error} is shown in (c). 
Tunnel currents amplitudes in this figure are presented without the account of the pair current suppression ($\alpha_{\rm supp}=1$).
}
\end{center}
\end{figure}

It is straightforward to generalize the microscopic model outlined above to a large Josephson junction of arbitrary 2D geometry. We have, for the
dynamics of the superconducting phase difference~$\varphi(\mathbf{r},t)$,
\begin{equation}
\frac{\p^2\varphi}{\p t^2} -\left(1+\beta\frac{\p}{\p t} \right) \nabla^2 \varphi + \alpha_N \frac{\p\varphi}{\p t} + \bar j(\mathbf{r},t) =0
\label{MM2D}
\end{equation}
\begin{equation}
\begin{split}
\bar{j}(\mathbf{r},t)=
\frac{k}{\Re \tilde{j}_p(0) }
\int_0^{\infty}\Big\{ 
j_p(kt')\,\sin\left[\frac{\varphi(\mathbf{r},t)+\varphi(\mathbf{r},t-t')}{2}\right] \\
+\, \bar{j}_{qp}(kt')\,\sin
\left [ \frac{\varphi(\mathbf{r},t)-\varphi(\mathbf{r},t-t')}{2}\right ] 
\Big\}\; dt',
\end{split}
\label{MM2D-jbar}
\end{equation}
where $\bar{j}(\mathbf{r},t)$ now plays a role of the current density (up to the subtracted normal current contribution), normalized to $V_g/A R_N$, where $A$ is the total area of the junction. The superconducting phase difference satisfies the Neumann-type boundary condition
\begin{equation}
\mathbf{n}\cdot\left(1+\beta\frac{\partial}{\partial t}\right)\nabla\varphi  = 
\mathbf{e}_z \cdot \left[\mathbf{n}\times\mathbf{h}  \right]
\end{equation}
where $\mathbf{n}$ is the in-plane outward normal and $\mathbf{h}$ is the normalized magnetic field in units $j_c \lambda_J$.

Even though the memory kernels $j_{p,qp}(\tau)$ allow explicit expression in terms of the Bessel functions (although, only at zero temperature~\cite{Harris-time-domain-1976}), the brute force approach to construct a finite difference scheme to solve the Eq.~\eqref{MM2D}
is struck with computational difficulties due to the need to evaluate the time integral~\eqref{MM2D-jbar} at each time step. This is especially not feasible in the case of large junction where such evaluation is needed at every node of the spatially discretized mesh. Therefore, an efficient algorithm to evaluate~\eqref{MM2D-jbar} is highly desirable.

Such algorithm was proposed by Odintsov, Semenov and Zorin~\cite{OSZ, OSZ-0} (OSZ).
Following this approach the time-domain kernels are fitted by a sum of complex exponentials,
\begin{equation}
\begin{split}
j_{p}(\tau)=\Re \sum_{n=0}^{N-1} A_n\, e^{p_n \tau}
\\
\bar{j}_{qp}(\tau)=\Re \sum_{n=0}^{N-1} B_n\, e^{p_n \tau}
\end{split}
\label{exps}
\end{equation}
where $A_n$, $B_n$ and $p_n$ ($\Re p_n < 0$) are complex parameters. 
Their values are obtained by fitting the tunnel current amplitudes in the frequency domain, $\tilde{j}_{p}(\xi)$ for the pair current and $\tilde{\bar{j}}_{qp}(\xi)=\tilde{j}_{qp}(\xi)-i\xi$ for the reduced quasiparticle current, by the Fourier transforms of the sums~\eqref{exps}, in accordance with the definition~\eqref{transforms}.
Introduction of the exponentials~\eqref{exps} allows to avoid the direct evaluation of the integral~\eqref{MM2D-jbar}. Substitution of~\eqref{exps} to~\eqref{MM2D-jbar} splits the integral into a finite number of composite parts whose values need only be
updated once per time step.

The first attempt to apply the OSZ algorithm to study dynamics of a long Josephson junction based on the MMT was made in Ref.~\onlinecite{GJ-1992}, however, with a limited success:
quantitative and qualitative disagreement of the numerical model from the analytical calculations were later realized~\cite{Hattel-1993} by the same authors. 
Unfortunately, based on the poor performance of their numerical model,
authors of Refs.~\onlinecite{GJ-1992, Hattel-1993} had drawn a conclusion about 
impossibility for the OSZ algorithm to reproduce essential characteristics of 
real Josephson junctions
and ceased their studies.
As we argue below, this conclusion was premature: in fact, the mediocre performance of the numerical model of Refs.~\onlinecite{GJ-1992, Hattel-1993} can be explained by the improper fit of tunnel current amplitudes in the subgap region. Furthermore, we show that with the use of~\eqref{exps}, the OSZ algorithm enables to achieve the MTT description of a Josephson junction, which is as good as if using the true kernels. Given that the true kernels are never known exactly, the minor difference between the two, if any, is irrelevant.

Our fit of tunnel current amplitudes by the expansion~\eqref{exps} with $N=8$ terms is presented in Fig.~\ref{fig:amps}a,b. The fit was obtained by calculating the complex parameters $p_n$, $A_n$, $B_n$ which minimize the cost function
\begin{equation}
\sum_X \int_0^2 D(X^{\rm fit},X^{\rm exact})^2\, d\xi,
\end{equation}
where 
\begin{equation}
D(X^{\rm fit},X^{\rm exact}) \equiv \frac{|X^{\rm fit}-X^{\rm exact}|}{\max(\tau_{a}/\tau_{r},|X^{\rm exact}|)},
\label{error}
\end{equation}
is the relative difference between the fitted and exact functions $X=\Re \tilde{j}_{p}(\xi)$, $\Im \tilde{j}_{p}(\xi)$, $\Re \tilde{j}_{qp}(\xi)$, $\Im \tilde{j}_{qp}(\xi)$, 
and $\tau_{a,r}$ are absolute and relative tolerances, respectively.
To stress a good performance of the obtained fit in the subgap region 
we redraw the imaginary parts of the tunnel current amplitudes by scaling them by a factor of~$20$: these are the curves which correspond to the vertical axis on the right in Figs.~\ref{fig:amps}a,b. 
As seen from the plot, the exact (dashed lines) and the fitted amplitudes (colored lines) are practically indistinguishable.
In order to make the comparison possible, we plot the relative difference defined by Eq.~\eqref{error} in Fig.~\ref{fig:amps}c. 
As seen from this figure, with $\tau_{a}/\tau_{r}=0.2$ we are able to achieve relative tolerance $\tau_{r}=0.005$ at an absolute tolerance $\tau_{a}=0.001$, which is sufficiently beyond the accuracy with which BCS tunnel current amplitudes can be relied on in description of real systems.
Finally, to convince ourselves that our own fit in Fig.~\ref{fig:amps} gives physically reasonable results consistent with that given by the true kernel functions, we carried out a numerical calculation for a benchmark model of a single fluxon used in Ref.~\onlinecite{Hattel-1993} and obtained an agreement between our analytical and numerical approaches.
Details of this calculation will be published elsewhere.
In practice, we have found that it has been always possible to reach a given precision by increasing the number of the fitting terms in the expansion~\eqref{exps}.
Therefore, the fit presented in Fig.~\ref{fig:amps} can be further improved, should the need arise 
 (for this, it is enough just to add exponentials with $\Im p_n$ in the regions of frequencies where the fit deviates the most).
It is, however, satisfactory enough for the purposes this fit is used for in the present paper.

Our numerical model with tunnel current amplitudes fitted by the 8 terms is only about 3 times slower than the conventional PSGE discretized by the same scheme. Given the complexity of the MTT, such a small difference between the MTT and PSGE may seem surprising and is explained as follows. The bottleneck of the numerical calculation with the PSGE is evaluation of a trigonometric function (the sine). In our numerical implementation of the MTT, only two such evaluations per time step are required, regardless of the number $N$ of the fitting exponentials. This gives a slow down by a factor of 2 plus some less significant $N$-dependent overhead. 
As a result, the performance of the numerical scheme is weakly dependent on the number of fitting exponentials.

To facilitate  evaluation of the quasiparticle and pair currents, and to motivate future theoretical studies of Josephson junctions based on the MTT, we designed C~code MiTMoJCo (Microscopic Tunneling Model for Josephson Contacts). MiTMoJCo is available as an open source under the GNU General Public License~\cite{mitmojco}
and can be used either in conjunction with available FEM and FDTD solvers or as part of a finite difference scheme in a standalone C code.

\section{Model of Josephson Flux Flow Oscillator}

An illustrative example of a Josephson system 
whose current-voltage characteristics can not be adequately described within the PSGE
is the Josephson flux flow oscillator (FFO)~\cite{Nagatsuma1,Nagatsuma2,Nagatsuma3,Nagatsuma4}. 
FFO is a long Josephson junction where a dense chain of fluxons driven by the electric current excites electromagnetic modes inside the junction.
To accommodate multiple fluxons and achieve a flux-flow regime, the length of the Josephson junction used as a FFO 
exceeds the Josephson penetration length by a large factor.
The potential of FFO for practical applications has been justified by development of a superconducting integrated receiver (SIR)~\cite{SIR, Koshelets-Shitov} which was successfully used in remote heterodyne spectroscopy of the Earth atmosphere on board of high-altitude balloon~\cite{Lange-2010,SIR-TELIS}, as well as first spectral measurements of THz radiation emitted from intrinsic Josephson junction stacks (BSCCO mesa) at frequencies up to 750 GHz~\cite{Li-2012,Koshelets-IEEE-2015}. 

To describe properties of FFO such as the linewidth and IVC,
all known theoretical studies of FFO rely on the PSGE (see, e.g., Refs.
~\onlinecite{Soriano, Golubov-FFO, Ustinov-chaos, Kurin-FFO, Cirillo, Salerno-FFO, 
Jaworski-1999, Yulin-FFO, Pankratov-2002-line, Pankratov-2002-driving, Sobolev-2006, Pankratov-2007,
Pankratov-JAP-2007, Khapaev-2008, Pankratov-PRB-2008, Pankratov-2008, Jaworski-2010, 
Matrozova-2011, Revin-2012, FFO-MCQTN}, to name a few). The most advanced of the FFO IVC models include a phenomenological
 modification of the damping parameter~\cite{Pankratov-2007,FFO-MCQTN}
 to reproduce the self-coupling effect manifested in the experimental IVCs~\cite{sc-Koshelets}.
In the microscopic model of FFO, which we introduce below, such modification is not necessary as the coupling of the junction to electromagnetic field comes naturally within the formalism of the MTT.
From the computational side, our numerical model of FFO
outperforms the voltage-dependent damping model~\cite{Pankratov-2007,FFO-MCQTN} as 
it is free from the iterative procedure needed in the voltage-damping model
to adjust the damping parameter,
rather, the DC component of voltage is obtained in a single run. Indeed, as our performance study shows, during one run of our simulation with the microscopic model, the voltage-dependent model would only be able to perform 3 iterations, which is far from being enough for the damping parameter to settle (typically, 20-30 iterations were required for convergence in Ref.~\onlinecite{FFO-MCQTN}).

Typically, the radiation generated by a FFO is used to drive a SIS mixer coupled via matching circuitry.
To improve impedance matching, the geometry of FFO is optimized by tapering off the width of a junction towards its ends. 
For realistic modeling of FFO, it is essential to take into account such variation of the junction width.
It is known~\cite{Benabdallah-1996} that the two-dimensional model for long Josephson junction with variable width can be reduced to a quasi one-dimensional model.
In a similar way, the quasi one-dimensional microscopic model of FFO 
derived from~\eqref{MM2D} takes a form,
\begin{widetext}
\begin{equation}
\varphi_{tt} + \alpha_N \varphi_t - \left(1+\beta\frac{\p}{\p t}\right)\varphi_{xx} 
- \frac{W'(x)}{W(x)}\left[h_{\rm ext} + \left(1+\beta\frac{\p}{\p t}\right)\varphi_{x}\right]
  + \bar j(x,t) -\Gamma_{\rm eff}(x) =0,
\label{FFO-MM}
\end{equation}
\begin{equation}
\begin{split}
\bar{j}(x,t)=
\frac{k}{\Re \tilde{j}_p(0) }
\int_0^{\infty}\Big\{ 
j_p(kt')\,\sin\left[\frac{\varphi(x,t)+\varphi(x,t-t')}{2}\right]
+\, \bar{j}_{qp}(kt')\,\sin
\left [ \frac{\varphi(x,t)-\varphi(x,t-t')}{2}\right ] 
\Big\}\; dt'
\end{split}
\label{FFO-jbar}
\end{equation}
where the $x$-dependent superconducting phase difference $\varphi(x,t)$ satisfies boundary conditions at the FFO's ends,
\begin{equation}
\varphi_x(- L/2,t) = -h_{\rm ext},\quad 
\varphi_x(L/2,t) + \beta \varphi_{xt}(L/2,t) = -h_{\rm ext} - \sigma(t).
\label{FFO-bc}
\end{equation}
\end{widetext}
Here, $L$ and $W(x)$ are the normalized length and width of the junction, respectively,
$\sigma(t)$ is the normalized electric current via the load in units $j_c\lambda_J W(L/2)$, and
$h_{\rm ext}$ is the normalized external magnetic field in units $j_c\lambda_J$. 
For an overlap junction geometry and, assuming an in-plane symmetry along the $x$ axis, we have for the effective bias current
\begin{equation}
\Gamma_{\rm eff}(x)=\frac{2 h_{\gamma}(x)}{W(x)}
\label{Gamma}
\end{equation}
where $h_{\gamma}(x)$ is the normalized magnetic field along the longest dimension of FFO, induced by the bias current. The two are related by the Maxwell equations which yield
\begin{equation}
2\int_{-L/2}^{L/2} h_\gamma(x) dx = \gamma \tilde{A}
\end{equation}
where $\tilde{A}\equiv A/\lambda_J^2$ is the normalized area of the junction and $\gamma$ is the bias current in units of the critical current~$A j_c$.
The magnetic field $h_\gamma(x)$ is related to the distribution of current in the electrodes feeding the FFO. Precise distribution of the magnetic field around the FFO should follow from the 
3D electromagnetic modeling with account of the leads, for example, using the available software~\cite{3D-MLSI, Khapaev-2008, Khapaev-2010, Khapaev-2015}.
Note, that the model~\eqref{Gamma} of FFO with tapered ends implies the rise of the effective bias current $\Gamma_{\rm eff}(x)$ towards the edges of the junction, which is not related to the electrodynamics of the junction but is merely a consequence of its geometry.
For sufficiently sharp ends and a linearly decreasing width $W(x)\sim \Delta x$ in proportion to the distance from the edges $\Delta x$, the rise $\Gamma_{\rm eff}(x)\sim 1/\Delta x$ can dominate the electrodynamic rise of the magnetic field $\sim 1/\sqrt{\Delta x}$ in a superconducting strip~\cite{sqrt}.
Despite a number of theoretical studies on the influence of an inhomogeneous bias current~\cite{Pankratov-2002-driving, Pankratov-2007, Jaworski-2008, Matrozova-2012},
and, given the developed theory of FFO with variable width~\cite{Benabdallah-1996, Benabdallah-2000, Carapella-2002, Jaworski-2005}, 
the effect of a purely geometrical rise of the effective bias current~\eqref{Gamma} on the IVC of a real FFO
seems to be largely ignored.

The model of FFO coupled to the RC load proposed in Ref.~\onlinecite{Soriano}
has been widely used in a number of subsequent theoretical studies~\cite{Sobolev-2006, Pankratov-2007, Pankratov-PRB-2008, Pankratov-APL-2008, Matrozova-2011, Revin-2012}.
However, the load impedance of a realistic system may be very different from the ideal case of a pure RC load.
Thus, a unified approach which enables to account for coupling of FFO to a an arbitrary load is highly desirable.
Assume one end of a FFO is coupled to a load described by a general impedance $Z(\omega)$. 
The time derivative of the superconducting phase difference at the FFO end is related to the load current $\sigma(t)$ by the convolution, 
\begin{equation}
\varphi_t(L/2,t) = \int_0^t z(t-t')\,\sigma(t') dt'
\label{load-model}
\end{equation}
where $z(\tau)$ is the impulse response~\cite{Varperian} defined by the Laplace (Fourier) transform of the frequency domain impedance $Z(\omega)$ normalized to the characteristic impedance at the radiation end of FFO,
\begin{equation}
Z_c\equiv \frac{\hbar \omega_J}{2 e j_c \lambda_J W(L/2)}
\label{Zc}
\end{equation}
Eq.~\eqref{load-model} should be solved alongside the integro-differential equation~\eqref{FFO-MM}. We employ the same approach for evolving Eq.~\eqref{load-model} 
as was used for solving the integro-differential equation~\eqref{FFO-MM}, that is,
we fit the impulse response by a series of exponentials in the same form as it was done for the tunnel current amplitudes~\eqref{exps},
\begin{equation}
z(\tau)=R_L\delta(\tau) + \Re \sum_{n=0}^{N_z-1} C_n\, e^{q_n \tau}
\label{z-tau}
\end{equation}
where $q_n$ and $C_n$ are complex parameters, and we separated explicitly the Ohmic contribution described by the normalized load resistance $R_L$.
At the end of the simulation, the power radiated by FFO can be calculated by taking the time average,
\begin{equation}
P_{FFO} = \frac{V_g^2}{4 Z_c k^2}\; \overline{ \sigma(t) \varphi_t(L/2,t) }.
\label{PFFO}
\end{equation}

The model~\eqref{load-model}-\eqref{PFFO} is general and can be applied 
to describe coupling of Josephson junction to an arbitrary load. A simple case of a load resistance $R_L$ and a single term with $q_0=0$, $C_0=1/C_L$ in the Eq.~\eqref{z-tau} corresponds to the model of RC load with parameters $R_L$ and $C_L$ used in Ref.~\onlinecite{Soriano} and the subsequent works~\cite{Sobolev-2006, Pankratov-2007, Pankratov-PRB-2008, Pankratov-APL-2008, Matrozova-2011, Revin-2012}. 
In this case, and with the assumption of the single harmonics at Josephson frequency dominating all other frequencies, Eq.~\eqref{PFFO} reduces to the Eq.(6) of Soriano~\cite{Soriano}.

\setlength{\unitlength}{0.1in}
\begin{figure*}
\begin{center}
		$
		\begin{array}{cc}				
		\begin{picture}(35,23)
		\put(0,0){\includegraphics[width=3.5in]{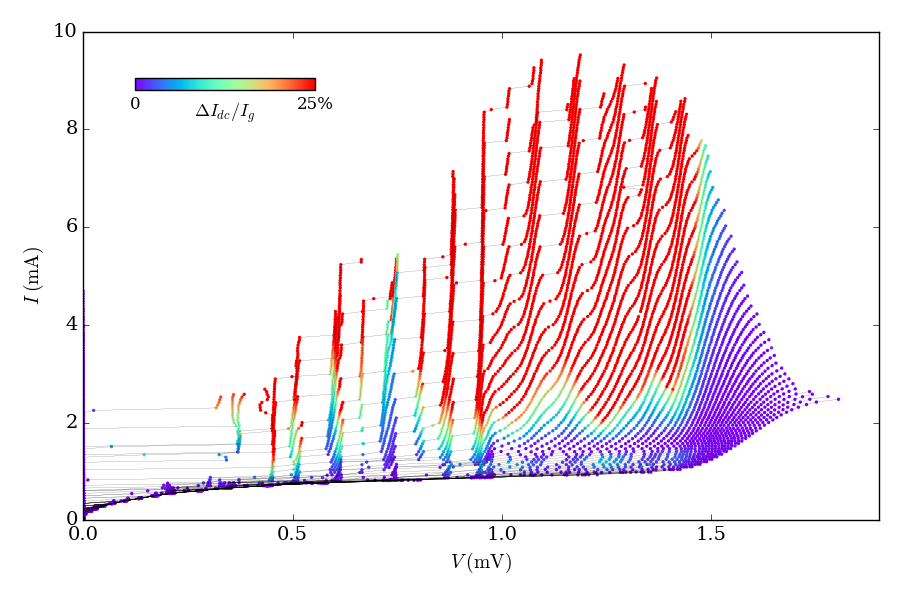}} 
		\put(31,20){(a)}
		\end{picture}	
		&
		\begin{picture}(35,23)
		\put(0,0){\includegraphics[width=3.5in]{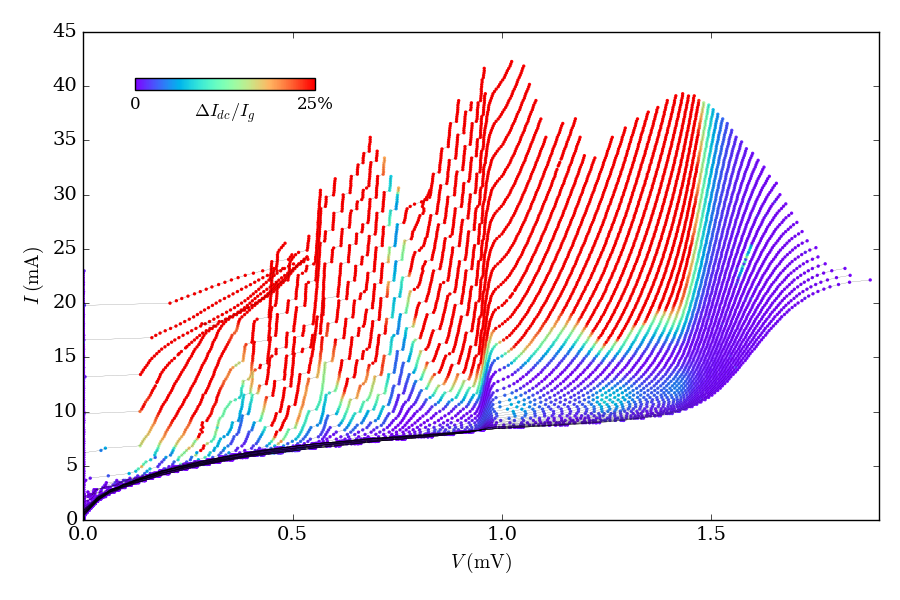}} 
		\put(31,20){(b)}
		\end{picture}	
		\\
		\begin{picture}(35,23)
		\put(-1,0){\includegraphics[width=3.6in]{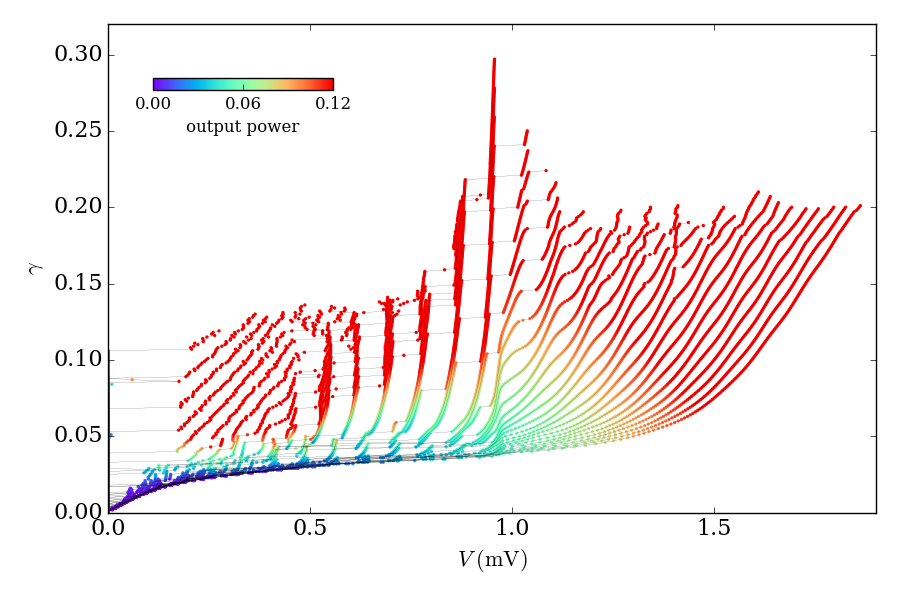}} 
		\put(31,20){(c)}
		\end{picture}	
		&
		\begin{picture}(35,23)
		\put(-1,0){\includegraphics[width=3.6in]{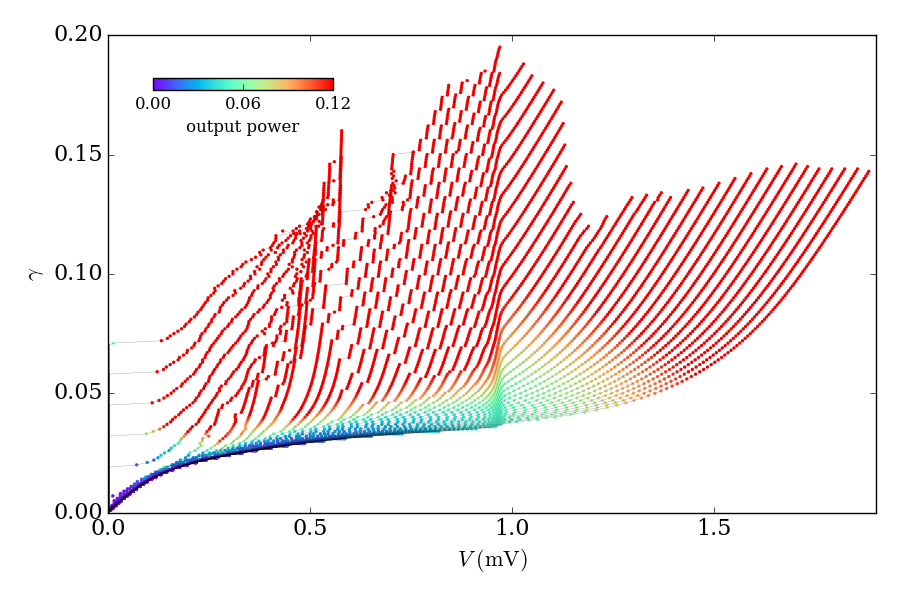}} 
		\put(31,20){(d)}
		\end{picture}					
		\end{array}
		$	
\caption{\label{ffo-iv} 
Experimental (a,b) and theoretical (c,d) IVCs of FFO. In experimental IVCs the color scale of its branches corresponds to the rise in the SIS mixer DC current from 0 to 25\% of the current step $I_g$ at the gap voltage (the more precise definition of $I_g$ is given in Ref.~\onlinecite{IRTECON}). 
The data for the SIS mixer where DC current rises above the 25\% threshold is painted by the same (red) color as the 25\% rise. Power output in the numerical IVCs is expressed in units $V_g^2/Z_c k^2$ and is cut at the 0.12 threshold.
(a) and (b) show experimental IVCs of the FFOs with length $80\rm\;\mu m$ and $400\rm\;\mu m$, respectively.
The corresponding numerical IVCs calculated with the use of the MTT are presented in (c) 
for $80\rm\;\mu m$ and (d) for $400\rm\;\mu m$. Values of the normalized external magnetic field $h_{\rm ext}$ vary with the step 0.07 from 1.20 to 4.28. In both numerical calculations Josephson penetration length is taken to be 5.5 $\rm\mu m$, normalized gap frequency $k=3.3$, surface damping $\beta=0.02$ and pair current suppression $\alpha_{\rm supp}=0.7$.
}
\end{center}
\end{figure*}
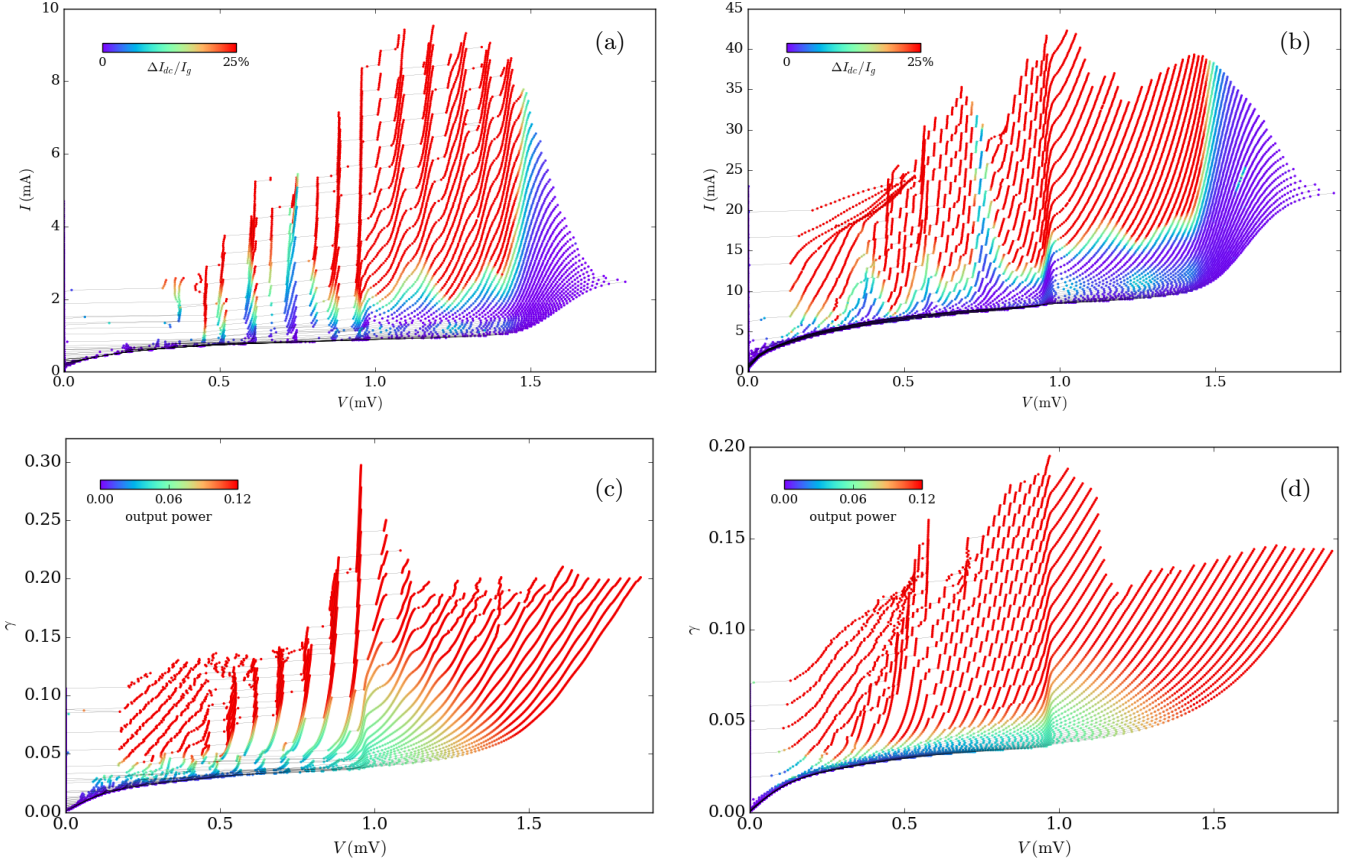

\section{Comparison to experimental results}

FFOs of lengths 60, 80, 120, 180, 250 and 400 $\rm\mu m$ were fabricated using the Nb-$\rm AlO_x$-Nb techonology.
The details of the fabrication process and design of the measurement system are similar to the previous experimental studies of FFO (see, e.g. Refs.~\onlinecite{Koshelets-Shitov,Koshelets-2004,Koshelets-2005,Koshelets-2007,Koshelets-IEEE-2015}). 
The layouts of the FFO samples were optimized for coupling to load by using the sharpened edge geometries:
the width 16 $\rm\mu m$ in the central region was degraded linearly to about 1 $\rm\mu m$
on a distance 40 $\rm\mu m$ (30 $\rm\mu m$ for the 60 $\rm\mu m$ junction) from either end.
The experimentally measured IVCs of two FFO samples of lengths 
$80\rm\;\mu m$ and $400\rm\;\mu m$ are shown in Figures~\ref{ffo-iv}a and b, respectively.
Each curve from the set of the shown IVC branches corresponds to a fixed value of the external magnetic field.
The color scale denotes a relative increase of the SIS DC current,
that is, 0 to 25\% rise compared to the height of the current step $I_g$ at the gap voltage
(the precise definition of $I_g$ is given in Ref.~\onlinecite{IRTECON}). 

We used the microscopic description of FFO introduced in the previous section for modeling the IVCs of the experimental samples. 
The differential part of the integro-differential equation~\eqref{FFO-MM} was discretized using the 2nd order central differences (for the derivatives $\varphi_{tt}$, $\varphi_{t}$, $\varphi_{xx}$ and $\varphi_{x}$), whereas 1st order discretization was used for the terms with surface damping $\beta$. Despite that the 1st order discretization introduces an error $O(\beta \Delta t)$ per time step, due to the smallness of $\beta$, the numerical scheme remains effectively 2nd order 
(compare the values $\beta=0.02$ for the surface damping and $\Delta t= 0.0227$ for the time step used in our calculations).
This enables to construct a semi-implicit scheme while having at hand a convenient explicit expression for the superconducting phase difference ready for the next step. 
Note, that in this scheme it is the normal resistance part of the quasiparticle current in Eq.~\eqref{reduced} that is taken into account implicitly, while the rest of tunnel current enters explicitly. 
In our calculations we assume a homogeneously fed FFO with $h_\gamma(x)=const$, which, according to~\eqref{Gamma}, results in an {\it increase} of the effective bias current towards the edges.
To account the coupling to load, a realistic load impedance was fitted
by three terms in Eq.~\eqref{z-tau} with parameters $C_0=C_1=C_2=0.5$, $q_0=-0.02$, $q_1=-0.02+1.1i$, $q_2=-0.01+3.75i$ and ${R_L=0.002}$. These values were estimated from results of our $\rm MathCad$ calculation for a real microwave circuit designed to provide wideband coupling of FFO to a SIS detector~\cite{Koshelets-2004,Koshelets-2007}.
According to~\eqref{z-tau}, the model impedance takes a form of a series of peaks at normalized angular frequencies $|{\rm Im}\,q_n|$ with characteristic widths defined by~$|{\rm Re}\,q_n|$.
We take $Z_c=1.6\rm\,\Omega$ as an estimate of the characteristic impedance given by the Eq.~\eqref{Zc}, Josephson penetration length $5.5$ $\rm\mu m$, normalized gap frequency $k=3.3$, and
surface damping parameter $\beta=0.02$. We took the pair current suppression parameter $\alpha_{\rm supp}=0.7$ as a reasonable estimate for the strong coupling correction for Nb junctions~\cite{Broom-IBM, Broom-IEEE} (the proximity effect~\cite{Golubov-1995, Golubov-1989} is expected to have a smaller effect in our junctions, not exceeding 10\%~\cite{Dmitriev}).
To improve computation of DC voltage we used the optimum filtration procedure for a sinusoidal signal, introduced in Ref.~\onlinecite{OSZ}.

The numerically calculated IVCs for the two experimental samples whose IVCs were shown in Figs.~\ref{ffo-iv}a,b,
are presented in Figs.~\ref{ffo-iv}c,d.
The color scale corresponds to the output power in units $V_g^2/Z_c k^2$ calculated using the Eq.~\eqref{PFFO} and are cut at the value 0.12 to match the 25\% saturation threshold as in the experimental IVCs.
Both in the experiments and the numerical calculations, the bias current rises from zero until reaching the end of the flux flow branch (in following, referred to as maximal flux flow current, MFFC). 
With further increasing the current beyond the MFFC value, 
the state of the junction switches from the flux flow regime to the ordinary phase rotation in the vicinity of the gap voltage.

The experimental and theoretical IVCs show a good overall agreement, although,
few features in which the two differ can be distinguished.
Firstly, the driving power of SIS mixer shown by the color in the experimental IVCs exhibits few peaks and deeps related to the frequency-dependent coupling between the FFO and SIS which is not taken into account in our theoretical model.
Nevertheless, our model does catch qualitatively the expected power output of the FFO in the region of small and moderate voltages.
At voltages above about $1.4\rm\;mV$ the theoretical model predicts a significantly higher power output in contrast to the experimental IVC where a sharp crossover to low SIS pumping is visible. This is attributed to the onset of damping in the experimental superconducting circuits when frequency of the FFO  
reaches the Nb gap frequency close to $700$ GHz. Note, that profiles of the experimental and theoretical IVC branches at $V>1.4\rm\;mV$ are also qualitatively different, which can be explained by influence of non-equilibrium effects. Indeed, in the region where the Josephson frequency exceeds the Nb gap frequency, splitting of Cooper pairs via absorption of electromagnetic quanta results in excess of quasiparticles. Such effects are not taken into account by the conventional MTT derived in the assumption of the equilibrium occupation of electron states. Development of the non-equilibrium MTT of Josephson tunnel junctions, therefore, would be highly beneficial for a complete theoretical description of Josephson FFO.

In general, shapes of the theoretical IVC curves match well that of the experimental ones: all of them exhibit a sharp crossover at the voltage $V_g/3$
due to an increase in the quasiparticle current 
and which is a direct manifestation of self-coupling~\cite{sc-Hasselberg, sc-Maezawa, sc-Koshelets}.
Both theoretical and experimental IVCs for 400 $\rm\mu m$ junction exhibit a definite cusp at about 1.2 mV where
the MFFCs of the IVC branches reach minimum. 
The effect seems to have a universal character for sufficiently long junctions and is exhibited 
also by FFOs of lengths 250 and ${180\;\rm\mu m}$.

The second feature, in which the theoretical and experimental IVCs differ, is 
that above the boundary voltage $V_g/3$ most of the theoretical IVC branches have smaller MFFCs as compared to the experimental curves.
A possible explanation could be the influence of the idle region~\cite{idle-Lee-1991, idle-Lee-1992, idle-Caputo-1994, idle-Monaco-1995, idle-Thyssen-1995, idle-Caputo-Int-1996, idle-Caputo-JAP-1999, idle-Franz-2001, 
idle-ZFS} which may have a stabilizing effect on the dynamics of FFO and, presumably, affect the values of MFFCs. 
Influence of the idle region on the dynamics of FFO has been neglected in our theoretical treatment (except for the renormalization of Josephson penetration length
on which it has an effect~\cite{idle-Caputo-Int-1996, idle-Caputo-JAP-1999}). 
The proper account of the idle region requires 
upgrading the model~\eqref{FFO-MM} to the full 2D problem~\eqref{MM2D} coupled to the Maxwell equations inside the idle region.
On the other hand, value of MFFCs may also be influenced by coupling to the load and affected by the losses in the matching circuitry. In a more advanced model of the coupling, the dynamics of the SIS junction and propagation of the electromagnetic waves with multiple reflections in the matching circuits should be solved simultaneously with~\eqref{MM2D}.
Due to the complexity of these factors, and, because of their dependence on specific details of the experimental setup, we leave this problem to future studies. 

It is interesting to note, that FFOs with small lengths exhibit Fiske steps even in the region of high voltages $V>V_g/3$ where these are normally suppressed in longer FFOs by the onset of damping.
Fiske steps are well pronounced for the 60 $\rm\mu m$ and 80 $\rm\mu m$ junctions and are marginally visible for the 120 $\rm\mu m$ junction.
In our numerical calculations the crossover is influenced by the surface damping $\beta$ and the pair current suppression parameter $\alpha_{\rm supp}$. Presence of the latter favors the quasiparticle current and thus increases the role of damping. From the Fiske step visibility crossover manifested for the FFO length of about 120 $\rm\mu m$, 
an upper limit on the surface damping can be estimated to be roughly 0.03 at $\alpha_{\rm supp}=0.7$ 
and $T=4.2$ K.
A smaller value $\beta=0.02$, used in Fig.~\ref{ffo-iv}, is obtained by fitting the IVC of the longest (400 $\rm\mu m$) junction in the Fiske region area ($V<V_g/3$). In our comparison of the experimental and theoretical IVC curves we find a tendency towards smaller $\beta$ in the small voltage region ($V<V_g/3$) and a larger $\beta$ in the high voltage region ($V>V_g/3$). Although, the observed tendency is within an error margin, and, furthermore, is subjected to the uncertainty in values of other parameters, if confirmed, this could indicate that the surface damping by itself can be frequency-dependent.

To conclude this section, the presented theoretical model of FFO lays fundamentals for modeling of a realistic FFO. The self-coupling effect observed in the experimental IVCs is caught naturally within the methodology of the MTT.
In fact, due to the important role played by coupling of tunnel currents and electromagnetic waves in the dynamics of superconducting phase difference,
it is evident that any realistic modeling of FFO should rely on the MTT.

\section{Discussion}

The presented microscopic approach can give a fresh look at the rich physics and variety of phenomena in large Josephson junctions. Apart from the example of the conventional FFO studied here, an admittedly incomplete list of the affected systems and phenomena includes detection and excitation of sub-terahertz sound by long Josephson junctions~\cite{Polzikova, FFO-sound}, Cherenkov~\cite{Cherenkov1, Cherenkov2} and
exponentially shaped~\cite{Benabdallah-1996, Benabdallah-2000, Carapella-2002, Jaworski-2005} FFOs,
transmission line intersections and networks~\cite{Nakajima-I,Nakajima-II,DGulevich-cloning,DGulevich-phenomena,Caputo-cloning,Sobirov-2016},
Josephson frequency comb generators~\cite{comb-Solinas-SciRep, comb-Solinas-JAP},
annular Josephson junction~\cite{Davidson-1985, Ustinov-1992} and its variations~\cite{DGulevich-collider, Monaco-ellipse, Monaco-annular}, linear~\cite{DGulevich-waveguides} and nonlinear~\cite{DGulevich-shape-PRL,DGulevich-shape-PRB} fluxon modes in 2D junctions, 
Josephson vortex qubits~\cite{Wallraff-2000, Kemp-2002, Shaju-2004, Price-PRB-2010}, pumps~\cite{DGulevich-pump} and ratchets~\cite{Goldobin-ratchet, Carapella-ratchet-PRB, Carapella-ratchet, Salerno-ratchet, Ustinov-ratchet, Goldobin-ratchet-2012, Goldobin-ratchet-2016}. 

To foster further research in this area, and, to enlarge the range of applications of the MTT, we created numerical library MiTMoJCo~\cite{mitmojco}.
Our theoretical results supported by a good agreement with the experimentally measured IVCs of several FFOs validate the use of MiTMoJCo in studies of other Josephson systems.

The described model naturally incorporates the phase dependent dissipation. This term has recently attracted particular attention because of the control it gives over quasiparticle relaxation in qubits. Understanding effects associated with quasiparticle tunneling is of crucial importance for developing superconducting qubits such as fluxonium~\cite{cos-Pop} as well as Majorana-based topologically protected qubits based on superconductor-semiconductor hybrid systems~\cite{Mourik, Higginbotham, Aasen, Higginbotham, Das, Rokhinson, Deng, Rainis, Beenakker, Finck, Nadj, Nadj-PRB, Albrecht}. Interestingly, the numerical approach to quasiparticle tunneling implemented here is not limited to description of superconducting systems, but may, in principle,
be applied to semiconductor superlattices~\cite{Esaki-Tsu, Wacker} where analogous photon-assisted tunneling effects arise in presence of bichromatic and polychromatic driving field~\cite{Timo-PRB-2008, Timo-2008, Timo-PRL-2009}.

\section*{Acknowledgments}

D.R.G. acknowledges support from the grant 3.8884.2017/8.9 of the Ministry of Education and Science of Russian Federation. V.P.K. acknowledges support from the grant no.~8168.2016.2 within the State Program for Support of Leading Scientific Schools and the Russian Foundation for Basic Research grant no.~17-52-12051.

\appendix*

\section{Tunnel Current Amplitudes}

For symmetric junction made of identical superconductors the normalized tunnel current amplitudes at $T=0$ are
\begin{equation}
\Re j_p(\xi)=
\begin{cases}
\, \frac{1}{2}\, K(\xi^{2}),\quad\quad |\xi|<1\\
\, \frac{1}{2|\xi|}\,K(\frac{1}{\xi^{2}}),\quad |\xi|>1
\end{cases}
\label{Rejp0}
\end{equation}

\begin{equation}
\Im j_p (\xi)=
\begin{cases}
0,\quad\quad |\xi|<1\\
-\frac{1}{2\xi}\, K\left(1-\frac{1}{\xi^{2}}\right),\quad |\xi|>1
\end{cases}
\label{Imjp0}
\end{equation}

\begin{equation}
\Re j_{qp} (\xi)=
\begin{cases}
\frac{1}{2}\, K(\xi^{2})- E(\xi^{2}),\quad\quad |\xi|<1\\
\left(|\xi|-\frac{1}{2|\xi|}\right)\,K\left(\frac{1}{\xi^{2}}\right)
-|\xi| E(\frac{1}{\xi^{2}}),\quad |\xi|>1
\end{cases}
\label{Rejqp0}
\end{equation}

\begin{equation}
\Im j_{qp}(\xi)=
\begin{cases}
0,\quad\quad |\xi|<1\\
\xi\, E\left(1-\frac{1}{\xi^{2}}\right)- \frac{1}{2\xi} K\left(1-\frac{1}{\xi^{2}}\right),\quad |\xi|>1
\end{cases}
\label{Imjqp0}
\end{equation}
where $\xi=\omega/\omega_g$ and $K$ , $E$ are complete elliptic integrals 
of the first and second kind correspondingly. Here we use the convention of elliptic functions taking {\it square} of the elliptic modules as an argument (note that Refs.~\onlinecite{Werthamer,Barone,Likharev} use a different convention for the elliptic integrals).

Current amplitudes at arbitrary temperature ${T\ge0}$ were given by Larkin and Ovchinnikov~\cite{Larkin}. 
For the Josephson junction formed by superconductors with gap energies $\delta_1\equiv \Delta_1/\omega_g$ and $\delta_2\equiv \Delta_2/\omega_g$ normalized to the gap frequency $\omega_g\equiv \Delta_1+\Delta_2$,

\begin{widetext}

\begin{equation}
\Re \tilde{j}_p (\xi)= \frac{\delta_1 \delta_2}{2} \int_{-\infty}^{\infty} 
\tanh\left(\alpha|\eta|\right)
\Big\{ \frac{\Theta(\delta_1-|\eta-\xi|)\,\Theta(|\eta|-\delta_2)}{\sqrt{\delta_1^2-(\eta-\xi)^2}\sqrt{\eta^2-\delta_2^2}} 
+ \frac{\Theta(|\eta|-\delta_1)\,\Theta(\delta_2-|\eta+\xi|)}{\sqrt{\eta^2-\delta_1^2}\sqrt{\delta_2^2-(\eta+\xi)^2}}\Big\}  d\eta
\label{Rejp}
\end{equation}

\begin{equation}
\Im \tilde{j}_p(\xi) = \, \frac{\delta_1\delta_2}{2}\int_{-\infty}^{\infty} 
\left\{\tanh\left[\alpha(\eta+\xi)\right] - \tanh \left(\alpha\eta\right) \right\}
\frac{\sgn(\eta)\, \sgn(\eta+\xi)\,\Theta(|\eta|-\delta_1)\,\Theta(|\eta+\xi|-\delta_2)}
{\sqrt{\eta^2-\delta_1^2}\sqrt{(\eta+\xi)^2-\delta_2^2}}d\eta
\label{Imjp}
\end{equation}

\begin{equation}
\Re \tilde{j}_{qp} (\xi) = -\, \frac12 \int_{-\infty}^{\infty} |\eta|\tanh(\alpha\eta)
\left[\frac{(\eta-\xi)\,\Theta(|\eta|-\delta_1)\,\Theta(\delta_2-|\eta-\xi|)}
{\sqrt{\eta^2-\delta_1^2}\sqrt{\delta_2^2-(\eta-\xi)^2}} 
+\frac{(\eta+\xi)\,\Theta(|\eta|-\delta_2)\,\Theta(\delta_1-|\eta+\xi|)}
{\sqrt{\eta^2-\delta_2^2}\sqrt{\delta_1^2-(\eta+\xi)^2}}
\right] d\eta
\label{Rejqp}
\end{equation}

\begin{equation}
\Im \tilde{j}_{qp}(\xi) = \frac12 \int_{-\infty}^{\infty} 
\left\{\tanh\left[\alpha(\eta+\xi)\right]-\tanh(\alpha\eta) \right\} 
\frac{|\eta| |\eta+\xi|\,\Theta(|\eta+\xi|-\delta_1)\,\Theta(|\eta|-\delta_2)}
{\sqrt{(\eta+\xi)^2-\delta_1^2}\sqrt{\eta^2-\delta_2^2}}\, d\eta
\label{Imjqp}
\end{equation}

\end{widetext}
where $\alpha\equiv \omega_g/2 k_B T$.
The correspondence to the original Larkin's~\cite{Larkin} expressions $I_{1,2,3,4}$ in their formula (22) is established by $\Re \tilde{j}_p (\xi)=I_1/\omega_g$, $\Im \tilde{j}_p (\xi)=I_2/\omega_g$, $\Re \tilde{j}_{qp} (\xi)=-I_4/\omega_g$, $\Im \tilde{j}_{qp} (\xi)=I_3/\omega_g$.
Note, that the original Larkin's expressions contains error in their formula for $I_1$
which was corrected here (see our note in Ref.~\onlinecite{misprints} for details). One may also check that
Eqs.~\eqref{Rejp}-\eqref{Imjqp} reduce to~\eqref{Rejp0}-\eqref{Imjqp0} in the zero temperature limit.

To obtain tunnel current amplitudes in Fig.~\ref{fig:amps} we assumed a symmetric junction ($\delta_1=\delta_2=1/2$) and smoothed the amplitudes by introducing 
a phenomenological peak width parameter $2\delta$ as described in Ref.~\onlinecite{Zorin}, 
\begin{widetext}
\begin{equation}
\Re \tilde{j}_{p,qp}(\xi) \to \Re \tilde{j}_{p,qp}(\xi)
-\frac{\xi\,\Re \tilde{j}_p(0)}{2\pi} 
\ln\left\{
\frac{\left[(1-\xi)^2+\delta^2\right] (1+\xi)^2}{(1-\xi)^2 \left[(1+\xi)^2+\delta^2\right]} 
\right\}
\end{equation}
\begin{multline}
\Im \tilde{j}_{p,qp}(\xi) \to \Im \tilde{j}_{p,qp}(\xi)
- \frac{\xi \alpha\, e^{\alpha}}{2\,(1+e^\alpha)^2}\; \ln\frac{\xi^2+\delta^2}{\xi^2}
\pm \frac{\xi\,\Re \tilde{j}_p(0)}{2} \Big[ \frac{2}{\pi}\arctan\frac{(1-\xi)}{\delta} -\sgn(1-\xi) \\
+ \frac{2}{\pi}\arctan\frac{(1+\xi)}{\delta}-\sgn(1+\xi) \Big]
\end{multline}
\end{widetext}
where the plus and minus signs in front of the square bracket 
in the last expression correspond to the pair and quasiparticle currents, respectively.
Parameter $\delta$ was estimated by comparing the smoothed $\Im \tilde{j}_{qp}(\xi)$ to 
the experimental IVC of voltage biased SIS mixer. We found that $\delta=0.008$ gives a good match to the measured mixer IVC. We used this value in calculation of tunnel current amplitudes in Fig.~\ref{fig:amps}.

Finally, the suppression of the pair current is taken into account by performing the replacement~\cite{Zorin-1983},
\begin{equation}
\tilde{j}_{p}(\xi) \to \alpha_{\rm supp}\, \tilde{j}_{p}(\xi).
\end{equation}


\begin{thebibliography}{99}
\bibitem{footnote1} The difference in definitions of $J_{p}(t)$ and $J_{qp}(t)$ can be removed by simply redefining $\tilde{J}_{p}\to \tilde{J}_{p}^*$ in Eq.~\eqref{WW}, however, we decide to keep the things as they are for historical reasons and consistency with earlier works.
\end{thebibliography}
\end{document}